\pdfoutput=1
\documentclass[a4paper,british,floatfix,pdftex,superscriptaddress,twocolumn,twoside,%
aps,prl,%
reprint,final% add final to see real layout, no todonotes anymore, ...
]{revtex4-2}%
\usepackage[T1]{fontenc}
\usepackage{graphicx}%
\usepackage[utf8]{inputenc}
\usepackage[todos]{CMImacros}

\hypersetup{citebordercolor=yellow,linkbordercolor=red,urlbordercolor=blue}

\newlength{\figwidth}
\newcommand*{\suppmeth}{Supplementary Methods\xspace}%
\newcommand*{\suppnote}[1]{Supplementary Note #1\xspace}%

\makeatletter
\def\@fnsymbol#1{\ensuremath{\ifcase#1\or \mathparagraph\or *\or \dagger\or \ddagger\or
      \mathsection\or \|\or **\or \dagger\dagger \or \ddagger\ddagger \else\@ctrerr\fi}}
\makeatother

\newcommand{\cfeldesy}{\affiliation{Center for Free-Electron Laser Science, Deutsches
      Elektronen-Synchrotron DESY, Notkestraße 85, 22607 Hamburg, Germany}}%
\newcommand{\uhhphys}{\affiliation{Department of Physics, Universität Hamburg, Luruper Chaussee 149,
      22761 Hamburg, Germany}}%
\newcommand{\uhhcui}{\affiliation{Center for Ultrafast Imaging, Universität of Hamburg, Luruper
      Chaussee 149, 22761 Hamburg, Germany}}%
\newcommand{\uhhchem}{\affiliation{Department of Chemistry, Universität Hamburg,
      Martin-Luther-King-Platz 6, 20146 Hamburg, Germany}}%

\newcommand{\mbi}{\affiliation{Max Born Institute, Max-Born-Straße 2a, 12489 Berlin, Germany}}%

\begin{document}
\title{Picosecond pulse-shaping for strong three-dimensional field-free alignment \mbox{of generic
      asymmetric-top molecules}}%
\author{Terry Mullins}\cfeldesy%
\author{Evangelos T.\ Karamatskos}\cfeldesy\uhhphys%
\author{Joss Wiese}\cfeldesy\uhhchem\uhhcui%
\author{Jolijn~Onvlee$^{\$}$}\cfeldesy\uhhcui%
\author{Arnaud~Rouz\'ee}\mbi%
\author{Andrey~Yachmenev}\cfeldesy\uhhcui%
\author{Sebastian Trippel}\cfeldesy\uhhcui%
\author{Jochen~Küpper$^{\ast,}$}\cfeldesy\uhhphys\uhhchem\uhhcui%
\date{\today}%
\begin{abstract}\noindent%
   *~Email: \href{mailto:jochen.kuepper@cfel.de}{jochen.kuepper@cfel.de};
   website:~\url{https://www.controlled-molecule-imaging.org} \\
   $^\$$~Current address: Institute for Molecules and Materials, Radboud University, Nijmegen, The Netherlands \\[1ex]
   {\centering\textbf{ABSTRACT}\\[1ex]} Fixing molecules in space is a crucial step for the imaging
   of molecular structure and dynamics. Here, we demonstrate three-dimensional (3D) field-free
   alignment of the prototypical asymmetric top molecule indole using elliptically polarized,
   shaped, off-resonant laser pulses. A truncated laser pulse is produced using a combination of
   extreme linear chirping and controlled phase and amplitude shaping using a
   spatial-light-modulator (SLM) based pulse shaper of a broadband laser pulse. The angular
   confinement is detected through velocity-map imaging of H$^+$ and
   C$^{2+}$ fragments resulting from strong-field ionization and Coulomb explosion of the aligned
   molecules by intense femtosecond laser pulses. The achieved three-dimensional alignment is
   characterized by comparing the result of ion-velocity-map measurements for different alignment
   directions and for different times during and after the alignment laser pulse to accurate
   computational results. The achieved strong three-dimensional field-free alignment of
   $\expectation{\cos^{2}\delta}=0.89$ demonstrates the feasibility of both, strong
   three-dimensional alignment of generic complex molecules and its quantitative characterization.
\end{abstract}
\maketitle%

\section{Introduction}
\label{sec:introduction}
Laser-induced alignment of gas-phase molecules has proven to be an efficient way to access the
molecular frame~\cite{Friedrich:PRL74:4623, Stapelfeldt:RMP75:543,Koch:RMP91:035005}. It was
extensively used in
high-harmonic-generation-spectroscopy~\cite{Torres:PRL98:203007,Weber:PRL111:263601},
strong-field-ionization~\cite{Pavicic:PRL98:243001, Meckel:NatPhys10:594,
   Trabattoni:NatComm11:2546}, x-ray-diffraction~\cite{Kuepper:PRL112:083002,
   Kierspel:JCP152:084307} and electron-diffraction~\cite{Meckel:Science320:1478,
   Hensley:PRL109:133202, Pullen:NatComm6:7262, Yang:NatComm7:11232, Walt:NatComm8:15651}
experiments, enabling the imaging of molecular structure and dynamics directly in the molecular
frame. Furthermore, it was crucial for retrieving the shapes of molecular
orbitals~\cite{Itatani:Nature432:867, Holmegaard:NatPhys6:428, Vozzi:NatPhys7:822}.

Such advanced imaging technologies are especially important for complex molecules, \ie, asymmetric
tops without any rotational symmetry, which is the case for almost all molecules on earth. Thus it
is of utmost importance to develop laser alignment into a practical tool for such molecules. This
would, for instance, maximize the information content of atomic-resolution imaging
experiments~\cite{Filsinger:PCCP13:2076, Barty:ARPC64:415}, as already suggested for the coherent
diffractive x-ray imaging of biological macromolecules more than fifteen years
ago~\cite{Spence:PRL92:198102}. In order to minimize perturbations by external fields this should be
achieved in a laser-field-free environment. The associated problems are twofold: The rotational
dynamics of these (generic) asymmetric top molecule molecules are very complicated and
incommensurate~\cite{Gordy:MWMolSpec,Thesing:JCP146:244304}. Moreover, the standard approaches to
characterize the 3D degree of alignment, using ion imaging of atomic fragments, mostly halogen
atoms, recoiling along a well-defined molecular axis, do not work.

One-dimensional alignment of linear and (near) symmetric top molecules has been demonstrated
extensively and really pushed to the limits~\cite{Friedrich:PRL74:4623, RoscaPruna:PRL87:153902,
   Stapelfeldt:RMP75:543, Holmegaard:NatPhys6:428, Trippel:PRA89:051401R, Trippel:PRL114:103003,
   Karamatskos:NatComm10:3364}, including concepts for time-domain detection methods for asymmetric
top molecules~\cite{Wang:PRA96:023424}. Furthermore, also the three-dimensional (3D) control of
rotation-symmetric molecules, typically during long laser pulses, was demonstrated by multiple
groups, making use of highly polarizable halogen atoms for large polarizability effects as well as
their symmetric fragmentation dynamics for characterization~\cite{Larsen:PRL85:2470,
   Tanji:PRA72:063401, Nevo:PCCP11:9912, Holmegaard:NatPhys6:428, Kierspel:JPB48:204002,
   Takei:PRA94:013401, Chatterley:JCP148:221105}. This was extended to the field-free 3D alignment
of asymmetric top molecules using sequences of either orthogonally polarized
~\cite{Underwood:PRL94:143002,Lee:PRL97:173001} or elliptically polarized laser
pulses~\cite{Ren:PRL112:173602}, as well as long-lasting field-free alignment in helium
nanodroplets~\cite{Chatterley:NatCommun10:133} using rapidly truncated pulses and the alignment of
one (generic) asymmetric top molecule molecule 6-chloropyridazine-3-carbonitrile using long laser
pulses~\cite{Hansen:JCP139:234313, Thesing:JCP146:244304}.

Here, we demonstrate and characterize the strong laser-field-free three-dimensional alignment of the
prototypical (bio)molecule indole (\ind, \autoref[a]{fig:indole}), a good representative of the
general class of molecules without any rotational symmetry and without any good leaving-group
fragments for standard characterization. We use a combination of a shaped, truncated,
elliptically-polarized laser pulse with a short kick pulse before truncation to induce strong
three-dimensional alignment. The degree of alignment is characterized through strong-field multiple
ionization and subsequent velocity-map imaging (VMI) of H$^+$, C$^+$, C$^{2+}$, and
CH$_{x}$N$^{+}$(x=0,1,2) fragments, combined with computational results to disentangle the temporal
and angular dependence of the alignment. Our approach shows that the molecular frame even of
generic asymmetric top molecules can be accessed.

\section{Results}
\label{sec:results}
\noindent\textbf{Experimental setup} \quad %
The experimental setup was described elsewhere~\cite{Trippel:MP111:1738}. Briefly, molecules were
cooled in a supersonic expansion from a pulsed Even-Lavie valve~\cite{Even:JCP112:8068}, operated at
a temperature of \celsius{80} and at a repetition rate of 100~Hz. Around 1.4~mbar of indole was
seeded in 95~bar of He, which was expanded into vacuum. The lowest-energy rotational states were
selected using an electrostatic deflector~\cite{Filsinger:JCP131:064309, Chang:IRPC34:557}. Inside a
VMI spectrometer, the strongly deflected molecules were aligned using a shaped 250~ps long laser
pulse with a peak intensity of $\sim1.25\times10^{12}~\text{W/cm}^2$. These pulses were produced by
a commercial laser system (Coherent Legend Elite Duo) with a $1$~kHz repetition rate and a spectrum
similar to a rounded saw tooth. The pulse was strongly negatively chirped to a duration of
$\ordsim600$~ps using a grating based compressor~\cite{Trippel:MP111:1738} before further shaping.
The alignment laser pulses were elliptically polarized with a $3:1$ intensity ratio between major
and minor axes.

The strongly chirped $9$~\mJ pulses were sent through a zero dispersion $4f$ pulse
shaper~\cite{Weiner:JQuantElec28:908} with a spatial light modulator (SLM, Jenoptik S640d) situated
at the Fourier plane in order to generate a truncated pulse with a fast fall-off time. The most
relevant part of the shaped temporal intensity profile, around the cutoff, is shown
in~\autoref[a]{fig:2DDOA}. The pulse consisted of a slow rise beginning at $-250$~ps (not shown),
followed by some amplitude modulation, a short kick with a duration of $2.6$~ps (FWHM), and,
finally, a fast truncation. The SLM was specifically used for spectral phase modulation of spectral
components between $815$~nm and $816$~nm. In addition, wavelengths longer than $\ordsim816$~nm were
blocked by a razor blade, situated directly in front of the SLM. This was necessary due to
Nyquist-sampling limits encountered. We note that the combination of phase shaping and spectral
truncation with the razor blade improved the temporal fall-off time by a factor of $2.5$ down to
$3.3$~ps, \ie, to within the noise level of the measurement, which was below $1$~\% of the signal
peak, compared to simply cutting the spectrum~\cite{Chatterley:JCP148:221105}.

The post-pulse observed at $\ordsim13$~ps is unwanted and probably originates from imperfect phase
compensation from the SLM or space-time coupling in the pulse shaping setup. However, the post pulse
is irrelevant to the degree of alignment within the first $10$~ps after the temporal truncation,
which corresponds to the important temporal region investigated in this experiment.

\begin{figure}
   % place here to not get it too early just for the reference to the indole structure...
   \includegraphics[width=\linewidth]{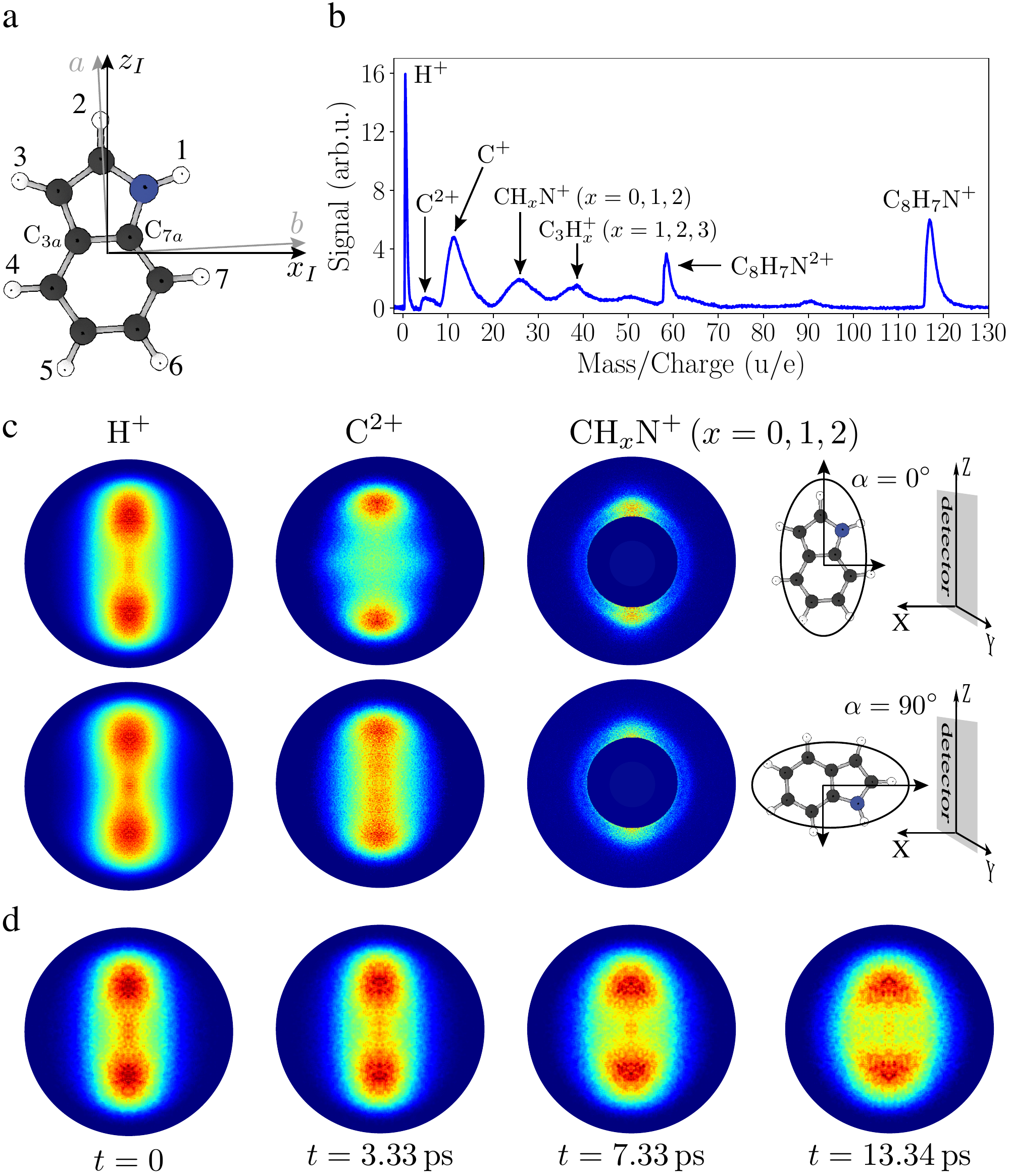}%
   \caption{\textbf{Molecular structure, mass spectrum, and ion images of indole.} (a) The structure
      of indole molecule with its principal axes of inertia and polarizability, labeled by $a,b,c$
      and $x_I,y_I,z_I$ ($\alpha_{y_I}<\alpha_{x_I}<\alpha_{z_I}$), respectively. (b) TOF mass
      spectrum of indole. (c) 2D momentum distributions for H$^{+}$, C$^{2+}$ and
      CH$_{x}$N$^{+}$(x=0,1,2) fragments at peak alignment at $t=3.3$~ps, with the major axis of the
      alignment laser polarization parallel (first row, $\alpha=0^{\circ}$) or perpendicular (second
      row, $\alpha=90^{\circ}$) to the detector plane. (d) Time-snapshots of 2D momentum
      distributions of H$^{+}$ fragments for the case of parallel laser polarization.}
   \label{fig:indole}
\end{figure}

A second, time-delayed, laser pulse with a pulse duration of 35~fs (FWHM) and a peak intensity of
$4.6\times10^{14}~\Wpcmcm$ was used to multiply ionize indole, resulting in Coulomb explosion. These
pulses were circularly polarized to avoid any secondary dynamics induced by electron
rescattering~\cite{Niikura:Nature417:917} and in order to minimize the bias from geometric
alignment.

Velocity-mapped fragments were detected on a microchannel plate (MCP) detector equipped with a
phosphor screen. The voltage on the MCP was switched between 2050~V (MCP ``on'') and 1150~V (MCP
``off'') using a fast switch (Behlke HTS 31-03-GSM) with $100$~ns rise- and fall-times to select the
different ion fragments based on their times of flight (TOF). A camera (Optronis CL600) recorded
single-shot images of the phosphor screen at $200$~Hz. Images without pulses from the molecular beam
were subtracted from those with the molecular beam to account for any signal from background
molecules in the interaction region. After selection of a suitable two-dimensional (2D) radial
range, the degree of alignment \cost was computed. A schematic visualization of the imaging geometry
is shown in~\autoref[c]{fig:indole}, with the detector plane defining the $(Y,Z)$ plane. The angle
$\theta_{\text{2D}}$ is defined as the polar angle in the detector plane with respect to the $Z$
axis. The angle $\alpha$ defines the orientation of the major polarization axis of the alignment
laser ellipse with respect to the lab-frame $Z$ axis, where $\alpha=0^{\circ}$ stands for parallel
alignment and $\alpha=90^{\circ}$ for perpendicular alignment, see~\autoref[c]{fig:indole}.

\medskip\noindent\textbf{Ion momentum distributions} \quad %
The in-plane principal axes of inertia ($a,b$) and polarizability ($z_I,x_I$) are shown in the
ball-and-stick representation of indole in \autoref[a]{fig:indole}. Both axis frames lie in the
plane of the molecule with an angle of \degree{2.75} between them, whereas the $c$ and $y_I$ axes
are perpendicular to that plane. The alignment process fixed the $z_I$ and $x_I$ axes in the
laboratory frame but not their directions, leading to four simultaneously present orientations of
the molecule. Upon Coulomb explosion, several fragmentation channels were detected. The resulting
time-of-flight mass spectrum is depicted in~\autoref[b]{fig:indole}.

Several ionic fragments showed anisotropic momentum distributions. The ion-momentum distributions of
H$^{+}$, C$^{2+}$ and CH$_{x}$N$^{+}\,(x=0,1,2)$ for a delay time of $t=3.3$~ps, corresponding to
the highest observed degree of alignment, are shown in \autoref[c]{fig:indole} for two orientations
of the alignment laser, \ie, with the main polarization axis being parallel, $\alpha=0^\circ$, or
perpendicular, $\alpha=90^\circ$, to the plane of the detector. $t=0$ corresponds to the peak
intensity of the alignment laser field. Furthermore, ion-momentum distributions of H$^{+}$ for time
delays of $t=0$, 3.3, 7.3, and 13.3~ps are shown in \autoref[d]{fig:indole}. The strongest
field-free alignment was observed near $t=3.3$~ps. At later delay times, the dephasing of the
rotational wavepacket leads to a decrease of the molecular alignment, as seen in the momentum
distributions recorded at time delays of $t=7.3~$ps and $t=13.3~$ps. Momentum distributions of other
fragments displaying alignment are shown in \suppnote{1}.

Indole does not contain unique markers, like halogen atoms, which would allow us to easily
experimentally access the degree of alignment. Therefore, all ions with a given mass to charge
ratio, produced through multiple ionization with subsequent Coulomb explosion, potentially
contributed to the measured 2D momentum distributions. There are seven sites in the indole molecule
from which the H$^{+}$ fragments originate and eight sites for the C$^{2+}$ fragments, see
\autoref[a]{fig:indole}. Each molecular site will result in different momentum and recoil axis of
the ionic fragment, and the total measured distribution is the sum of all of them.

\begin{figure}[t!]
   \includegraphics[width=\linewidth]{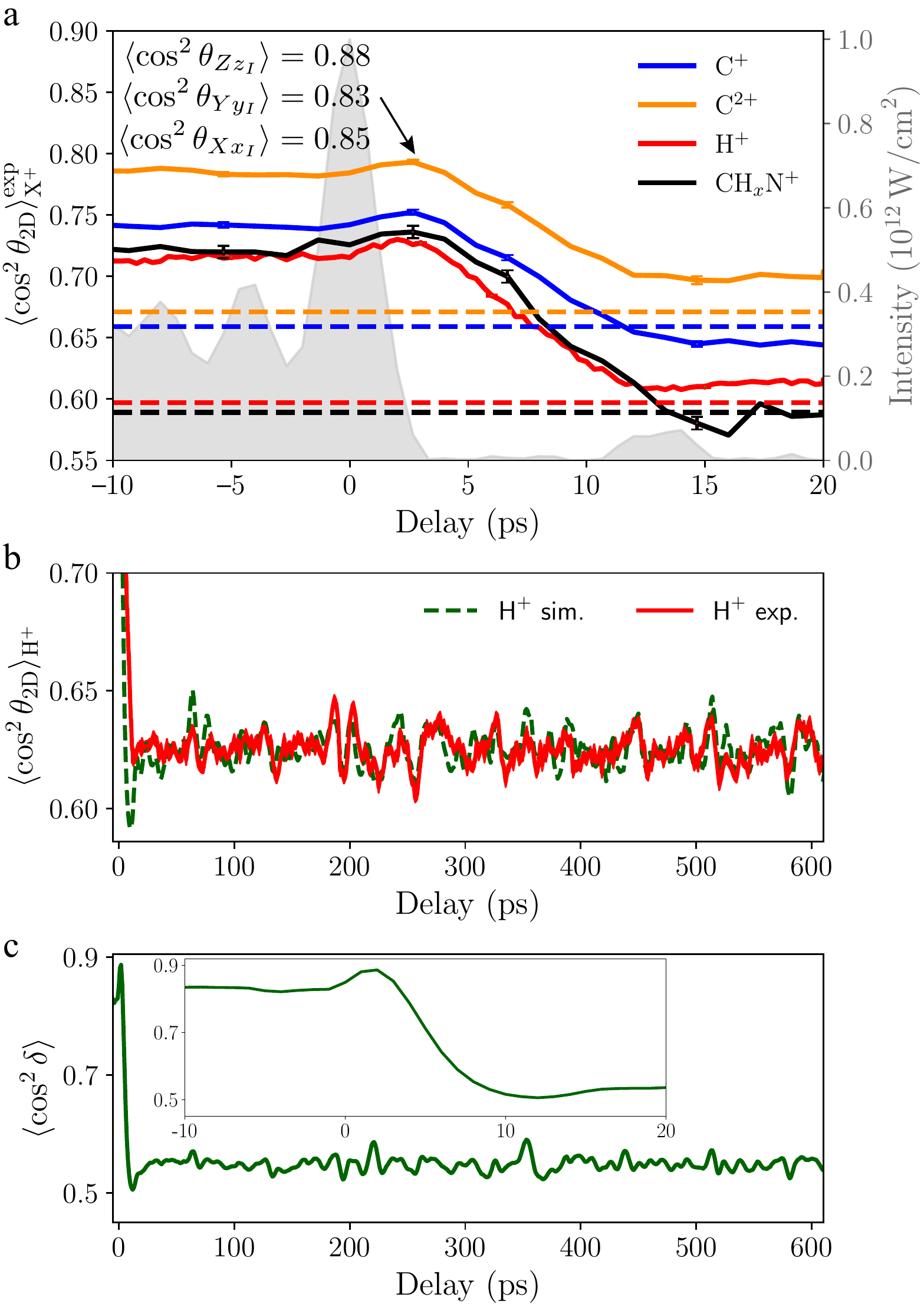}%
   \caption{\textbf{Temporal evolution of the alignment of indole for the parallel alignment
         geometry ($\alpha=0^{\circ}$).} (a) The solid lines show the measured 2D degree of
      alignment $\cost^\text{exp}_{\text{X}^+}$ for different fragments $\text{X}^+$ and the dashed
      lines indicate values of the 2D degree of alignment obtained without alignment laser.
      Statistical error bars, representing the standard error, are shown for selected delays. The
      grey area shows the intensity profile of the alignment laser pulse. Also shown are the three
      expectation values $\expectation{\cos^2\theta_{Xx_I}}$, $\expectation{\cos^2\theta_{Yy_I}}$
      and $\expectation{\cos^2\theta_{Zz_I}}$, with $\theta_{Ij}$ being the angles formed between
      the three main polarizability axes $j \in (x_{I},y_{I},z_{I})$ of indole and the three
      lab-frame axes $I \in (X,Y,Z)$, and their computed values for the actual 3D alignment of
      indole. (b) The alignment revival structure of H$^{+}$ fragments for longer times is shown in
      red, with the line thickness corresponding to the experimental standard error of the
      measurements. The dotted green line shows the fitted simulation for the H$^{+}$ fragment. (c)
      Simulated 3D degree of alignment, characterized through the single-scalar metric
      $\expectation{\cos^{2}{\delta}}$, see text/SI for details. Note the peak after truncation
      reaching $\expectation{\cos^{2}{\delta}}=0.89$ at $t=3.3$~ps.}
   \label{fig:2DDOA}
\end{figure}
The delay-dependent measured 2D degree of alignment is shown for a variety of fragments
in~\autoref{fig:2DDOA}. Assuming axial recoil, H$^{+}$ fragments would have measurable momentum
components only within the $ab$ plane of indole~\cite{Kukk:PRA99:023411}. Hence, the H$^{+}$
fragments are \emph{a priori} a good measure of the planar alignment of indole in the laboratory
frame. The slow rise of the alignment pulse confined the plane of the indole molecules in a
quasi-adiabatic fashion~\cite{Trippel:PRA89:051401R,Trippel:MP111:1738} to a measured maximum degree
of alignment of $\cost^\text{exp}_{\text{H}^+}=0.72$. Following the kick at the end of the alignment
pulse, the degree of alignment increased slightly to $\cost^\text{exp}_{\text{H}^+}=0.73$ before
monotonically decreasing over $\ordsim10$~ps to $\cost^\text{exp}_{\text{H}^+}=0.62$. The permanent
alignment of $\cost^\text{exp}_{\text{H}^+}=0.62$ was slightly higher than the value
$\cost^\text{exp}_{\text{H}^+}=0.60$ observed without alignment laser; the latter is due to the
geometric alignment from an isotropic distribution. At a delay of $3.3$~ps the intensity of the
alignment pulse decreased to 1~\% of its maximum, and the ``field-free'' region began. At this delay
the degree of alignment was $\cost^\text{exp}_{\text{H}^+}=0.73$, which was even larger than the
alignment measured just before the kick, confirming that the planar alignment in the field-free
region was even better than for an adiabatic alignment
pulse~\cite{Guerin:PRA77:041404,Rouzee:JPB41:074002}. All other fragments showed similar
distributions to the H$^{+}$ fragment, with the measured maximum degree of alignment being largest
for the C$^{2+}$ fragment. The differences in the measured alignment between the H$^{+}$, C$^{+}$,
C$^{2+}$ and CH$_{x}$N$^{+}$(x=0,1,2) fragments can be attributed to non-axial recoil or to the
geometry of Coulomb explosion fragmentation, \ie, the velocity vectors of the fragments in the
molecular frame.

\medskip\noindent\textbf{Characterization of 3D alignment} \quad %
To determine the 3D alignment of indole, an additional observable is required that characterizes the
in-plane alignment, \ie, the alignment of the most polarizable axis of indole $z_{I}$ with respect
to the main polarization axis of the alignment field. This information can be accessed by measuring
the angular distribution of the ionic fragments within the indole plane. By rotating the
polarization ellipse of the alignment laser around the laser propagation axis at a fixed delay time
of $3.3$~ps, the laboratory axes to which the $a$ and $b$ axes of indole align, were commensurately
rotated. In the laboratory frame, the transverse momenta of ionic fragments recoiling within the
plane of indole will depend on the ellipse-rotation angle $\alpha$, between $\alpha=0$ for parallel
and $\alpha=\degree{90}$ for perpendicular orientation, see \autoref[c]{fig:indole}. By counting
only those fragments impinging at the center of the detector, within a small radius of 20 pixel, the
distribution of fragments within the plane can be determined~\cite{Villeneuve:APB74:S157}. Note that
the size of the VMI images, as shown in \autoref[c]{fig:indole} and \autoref[d]{fig:indole}, was
$480\times 480$ pixels. Full tomographic measurements were carried out for H$^{+}$, C$^{+}$,
C$^{2+}$ and CH$_{x}$N$^{+}$(x=0,1,2) at $t=3.3$~ps for
$\alpha=\degree{0\text{--}180}~(\Delta\alpha=\degree{2})$. A visualization of the 3D reconstructed
signal is shown for the H$^{+}$ fragment in \suppnote{2}. For both, the H$^+$ and the C$^{2+}$
fragments, the 3D velocity distributions were quasitoroidal, \ie, no considerable density at or
around the origin was observed. The signals, measured at the center of the VMI in the 2D data, can
thus be attributed to in-plane fragments recoiling along the detector normal, proving the validity
of the approach chosen. The approach itself is equivalent to using narrow slices through the fully
reconstructed 3D momentum distributions from tomographic measurements, however it presents certain
advantages. These are in particular a decreased data acquisition time, since less data for the
characterization of the angular distributions are required than for a full tomographic
reconstruction, and finally the actual tomographic reconstruction of the 3D momentum distributions
can be circumvented, rendering the chosen approach more practical.

\begin{figure}
   \includegraphics[width=\linewidth]{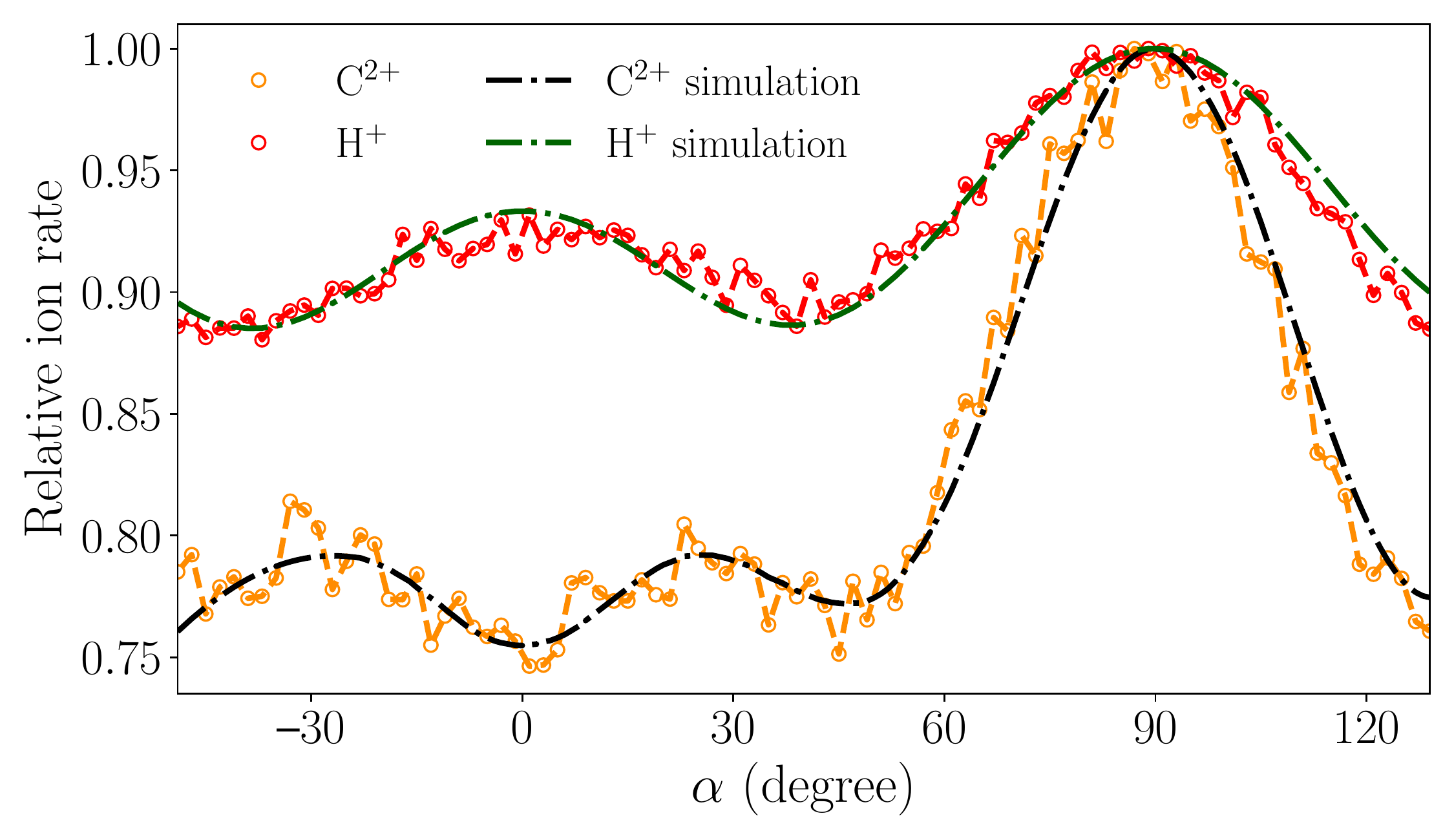}
   \caption{\textbf{In-plane angle-dependent ion count rates.} Measured masked-VMI ion count rates
      of H$^{+}$ and C$^{2+}$ fragments as the major axis of the alignment laser's polarization
      ellipse is rotated; see text for details. Circles (red and orange) indicate measured values,
      the dashed lines (green and black) are simulations based on fitted atomic-ion contributions,
      see text for details.}
   \label{fig:CPP_tom}
\end{figure}
Angular scans of this ``masked VMI'' are shown in \autoref{fig:CPP_tom} for H$^{+}$ and C$^{2+}$.
For both fragments, we observed a clear angle-dependent structure on top of a significant isotropic
background. C$^{2+}$ ions show two smaller peaks at $\alpha\approx\pm\degree{30}$ and a much
stronger peak at $\alpha\approx\degree{90}$. The H$^{+}$ signal shows a peak at
$\alpha\approx\degree{90}$, similar to C$^{2+}$, and a smaller peak at $\alpha\approx\degree{0}$.
Note that at \degree{90} the alignment laser's major polarization axis is pointing along the
detector normal. A direct extraction of the in-plane degree of alignment from the experimentally
obtained in-plane angular distribution was not possible due to a large isotropic background.
Furthermore, the degree of molecular alignment retrieved from the angular momentum distributions of
H$^+$ and C$^{2+}$ can be misrepresented mainly due to two reasons -- the many-body Coulomb break-up
of the multiply charged indole cation, with ionic fragments violating the axial-recoil
approximation, as well as the indistinguishability of fragments emitted at different molecular
sites.

\medskip\noindent\textbf{Computational calibration of the degree of alignment} \quad %
In order to determine the actual 3D degree of alignment, we performed comprehensive variational
simulations of the rotational dynamics of indole in the presence of the alignment field. We employed
the general variational approach RichMol~\cite{Owens:JCP148:124102, Yachmenev:JCP151:244118} to
compute time-dependent rotational probability density distributions for different delay times. In
order to incorporate the experimental conditions and to achieve better agreement, we took into
account the non-thermal distribution of rotational states in the deflected part of the molecular
beam and laser focal-volume averaging.

The total probability density distributions of C$^{2+}$ and H$^{+}$ were modeled as the weighted
sums of contributions from the individual atoms. As a consequence of orientational averaging, most
of the arising ions do not possess a unique recoil direction within the polarization frame of the
alignment laser. We accounted for this by using equal weights for pairs of atoms H1 \& H3, H4 \& H7,
H5 \& H6, C3a \& C7a, C4 \& C7, and C5 \& C6. As the recoil axes we choose vectors connecting carbon
atoms to the center of mass of the molecule for C$^{2+}$ and vectors along molecular C--H and N--H
bonds for H$^+$. To reproduce the experimental data the simple axial recoil approximation yielded
excellent agreement for C$^{2+}$, whereas for H$^+$ we had to account for non-axial recoil by
convoluting the calculated probability density distributions of hydrogen atoms with a Gaussian
function of a solid angle representing angular displacement from the recoil vector. The weights and
the FWHM parameter of this Gaussian function were determined in a least-squares fitting procedure to
the measured alignment revival trace and the angle-dependent masked VMI data. The obtained
parameters are specified in \suppnote{3}. The results of the fit show very good agreement with the
experimental alignment revivals in \autoref[b]{fig:2DDOA} and excellent agreement with the
integrated in-plane angle-dependent projections through the 3D momentum distribution in
\autoref{fig:CPP_tom}.

This excellent agreement confirms the correct representation of the experiment by our quantum
simulations. In principle, experimental input parameters to the simulations could be varied, but
this was not necessary due to an accurate determination of experimental parameters. Such a fitting
procedure is also undesirable, due to the time-consuming nature of the simulations. The obtained
planar alignment in terms of squared direction cosines~\cite{Makhija:PhysRevA85:033425} at
$t=3.3$~ps is $\expectation{\cos^2\theta_{Yy_I}}=0.83$ and $\expectation{\cos^2\theta_{Xx_I}}=0.85$.
These values are higher than the measured values, which is due to non-axial recoil of the H$^+$
fragments and different recoil axes contributing to the measured ion-momentum distributions. The
computed in-plane degree of alignment at $t=3.3$~ps is
$\expectation{\cos^{2}{\theta_{Zz_{I}}}}= 0.88$, which we also assign as the experimental value due
to the excellent match of the angular distributions in \autoref{fig:CPP_tom}. Simulated
time-dependent alignment revivals can be found in \suppnote{3}. A single scalar metric describing
the overall degree of 3D alignment
$\cos^2\delta=\frac{1}{4}(1+\cos^2\theta_{Zz_I}+\cos^2\theta_{Yy_I}+\cos^2\theta_{Xx_I})$~\cite{Makhija:PhysRevA85:033425},
is shown in~\autoref[c]{fig:2DDOA}. A maximum degree of field-free alignment of
$\expectation{\cos^{2}{\delta}}=0.89$ was obtained, which is comparable to or even larger than one
can achieve for complex asymmetric top molecules using adiabatic alignment
techniques~\cite{Hansen:JCP139:234313, Trippel:JCP148:101103} and clearly sufficient for
molecular-frame coherent diffractive imaging~\cite{Filsinger:PCCP13:2076, Barty:ARPC64:415}.

\section{Discussion}
\label{sec:discussion}
We demonstrated strong laser-field-free 3D alignment of the prototypical complex (generic)
asymmetric top molecule indole induced by shaped truncated quasi-adiabatic laser pulses. Both, the
amplitude and the phase of a strongly-chirped broad-band alignment laser pulse were tailored using
an SLM, which allowed us to produce very short truncation times, unachievable with amplitude
truncation alone. The combination of quasi-adiabatic alignment with a kick pulse directly before the
sudden truncation produced a higher degree of alignment under field-free conditions than in the
field. The already achieved strong degree of alignment is limited by the initially populated states
in the molecular beam~\cite{Kumarappan:JCP125:194309} and could be further improved through even
colder molecular beams~\cite{Chang:IRPC34:557, Filsinger:JCP131:064309}.

We have developed and tested a versatile approach to characterize molecular alignment in 3D by
performing separate measurements of planar alignment and in-plane tomography. Planar alignment was
characterized as the time-dependent alignment trace of the H$^+$ fragments. In-plane alignment was
characterized using the angular dependence of H$^+$ and C$^{2+}$ fragment distributions at the
center of the detector, obtained by rotating the laser polarization ellipse and thus the molecule in
the plane perpendicular to the detector. Robust variational simulations of the alignment dynamics of
indole, considering weighted contributions of fragments emitted non-axially at different molecular
sites, reproduced the experiment with high accuracy.

This demonstration of strong field-free alignment for an asymmetric top rotor without rotational
symmetries and without any good ionic fragments for the characterization of the alignment paves the
way for strong field-free alignment of any arbitrary molecule. This opens up important prospects for
probing native (bio)molecules in the molecular frame~\cite{Spence:PRL92:198102, Barty:ARPC64:415,
   Teschmit:ACIE57:13775} without chemically attaching marker atoms that influence the function and
properties of the molecule.

\section{Methods}
\label{sec:methods}
Our general experimental setup was described previously~\cite{Trippel:MP111:1738, Chang:IRPC34:557}
and the specific details for the current experiment were provided in the main text. Software used
for the simulations were described elsewhere~\cite{Owens:JCP148:124102, Yachmenev:JCP151:244118} and
specific details were described in the main text and the \suppmeth.

\subsection{Data availability}
The data that support the findings of this study are available from the repository at
\url{https://doi.org/10.5281/zenodo.5897172}.

\subsection{Code availability}
Quantum rotational dynamics simulations were performed using Richmol, available at
\url{https://github.com/CFEL-CMI/richmol}. Further codes used for analysis of experimental data and
analysis are available at \url{https://doi.org/10.5281/zenodo.5897172}.

\bigskip

\section*{Acknowledgements}
We thank Stefanie Kerbstadt for helpful discussions. We acknowledge support by Deutsches
Elektronen-Synchrotron DESY, a member of the Helmholtz Association (HGF), and the use of the Maxwell
computational resources operated at Deutsches Elektronen-Synchrotron DESY. This work has been
supported by the Deutsche Forschungsgemeinschaft (DFG) through the priority program ``Quantum
Dynamics in Tailored Intense Fields'' (QUTIF, SPP1840, KU 1527/3, AR 4577/4, YA 610/1; J.K., A.R.,
A.Y.) and by the Clusters of Excellence ``Center for Ultrafast Imaging'' (CUI, EXC~1074,
ID~194651731; J.K.) and ``Advanced Imaging of Matter'' (AIM, EXC~2056, ID~390715994, J.K.), and by
the European Research Council under the European Union's Seventh Framework Programme (FP7/2007-2013)
through the Consolidator Grant COMOTION (614507; J.K.). J.O.\ gratefully acknowledges a fellowship
of the Alexander von Humboldt Foundation.

\section*{Author contributions}
J.K.\ and A.R.\ devised the study, T.M., J.W., J.O., and S.T.\ set up the experiment, and T.M.,
J.W., and J.O.\ carried out the experiment. A.Y.\ and E.K.\ developed the simulations and performed
the calculations. T.M., J.W., J.O., and E.T.\ analysed the experimental and computational data.
J.K.\ supervised the study. All authors were involved in interpreting the data and clarifying
concepts in the experiment and its analysis. T.M.\ and E.K.\ wrote the initial manuscript and all
authors were involved in discussing and editing the manuscript.

\section{Supplementary Methods: \\[1ex] Experimental Details}
\label{sec:exp_details}
The peak intensity was determined by combining measurements of the pulse energy, the temporal
profile, and the spatial beam profile. The pulse energy was determined by measuring the average
power (Coherent PM30 power meter) and dividing by the repetition rate of $1$~kHz. The temporal
profile of the alignment pulse was determined experimentally by measuring a cross-correlation
between the alignment pulse and the Coulomb explosion pulse. Finally, the spatial beam profile was
measured on a beam profiler (Ophir Photonics, Spiricon SP620U). We note that the pulse shaper
affected both the temporal profile and the transversal spatial distribution of the beam, leading to
the so-called space-time coupling. As a result, the temporal profile was not homogeneous spatially
and the maximum degree of molecular alignment was achieved experimentally for a probe laser beam
that was spatially offset with respect to the alignment laser beam. To take this effect into
account, the spectrum of the alignment laser pulse was filtered using a $1$~nm band-pass filter
centered at a wavelength of $815$~nm. This section of the spectrum was chosen as it provided the
highest contribution to the intense peak observed in the temporal intensity profile of the alignment
laser pulse. We note that the peak intensity of the band-passed alignment laser pulse coincided with
the position of the probe laser beam that provided an optimal molecular alignment. The position of
this peak was used, in conjunction with the integrated spatial beam profile of the alignment laser,
\ie, including all wavelengths, to scale the measured integrated energy, which was only measurable
for all combined wavelengths. The scaling factor used, $I_{\mathrm{sc}}$, was defined as the ratio
between the intensity measured at the position of the probe laser where maximum alignment was
achieved and the peak intensity of the laser beam obtained without filtering. The peak intensity of
the alignment laser pulse that is shown in Fig. 2 of the main manuscript was then obtained using the
following expression:
\begin{equation}
   I_{0} = E \, I_\mathrm{sc} / \left(\int I_s(x,y) dx dy \int I_t(t) dt\right)
\end{equation}
with the measured pulse energy $E$, the normalized spatial intensity profile $I_s(x,y)$ obtained
from the beam profiler measurement, and the normalized temporal intensity profile $I_t(t)$ retrieved
from the cross-correlation measurement.

The statistical and systematic error of the peak intensity was estimated around $\ordsim7$~\% and
$\ordsim10$~\%, respectively. We note that the degree of alignment was not significantly changing
when the peak intensity was varied by $\pm10$~\%, in agreement with our error estimates. This was
also confirmed by simulations carried out for different peak intensities.

\section{Supplementary Note 1: \\[1ex] Ion-momentum distributions from strong-field ionization of aligned indole}
\label{sec:VMI_images}
Ion-momentum distributions for H$^{+}$, C$^{2+}$ and CH$_{x}$N$^{+}$(x=0,1,2) fragments recorded at
a time delay of $t=3.3$~ps, \ie, at the highest degree of field-free alignment, are shown in
Figure~1 of the main article.
\begin{figure}[bh!]
   \centering%
   \includegraphics[width=\linewidth]{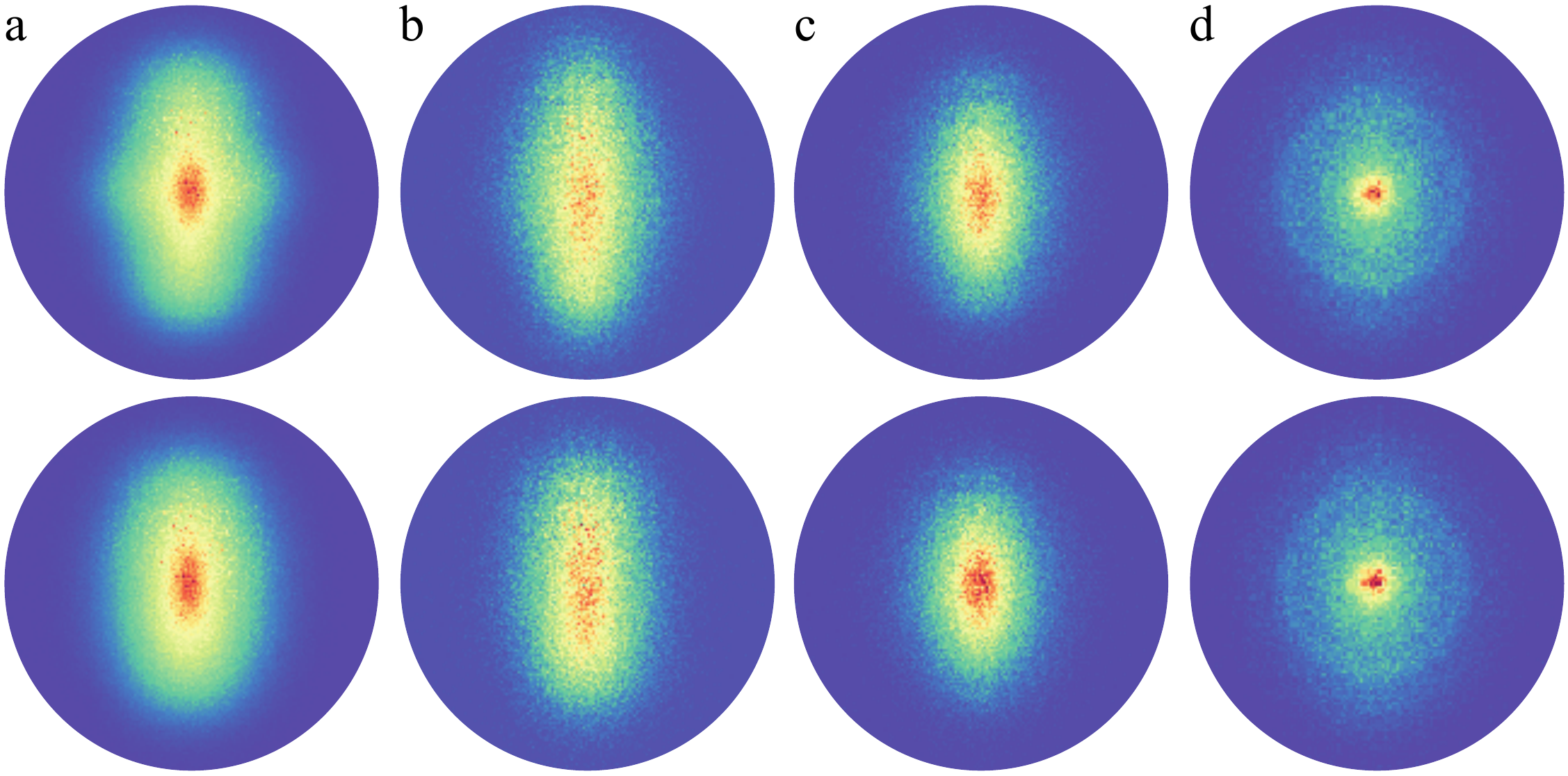}%
   \caption{\textbf{2D ion-momentum distributions} of the fragments (a) C$^+$, (b) C$_2^+$, (c)
      C$_3$H$_x^+$, and (d) C$_4$H$_x^+$. The top row shows images obtained for the major
      alignment-laser-polarization axis vertical and parallel to the detector surface whereas the
      bottom row shows corresponding images with the minor alignment-laser-polarization axis
      vertical and parallel to the detector surface; \cf Fig.~1 in the main article.}
   \label{fig:fragment_alignment}
\end{figure}
In addition, C$^+$, C$_2^+$, C$_3$H$_x^+$, and C$_4$H$_x^+$ ion-momentum distributions were also
recorded in the same experimental conditions and are displayed in \autoref{fig:fragment_alignment}.
For C$_3$H$_x^+$ and C$_4$H$_x^+$ fragments, $x$ corresponds to fragments with different number of
hydrogens whose masses could not be resolved by the high-voltage gating of the detector. In the top
row, ion-momentum distributions are shown with the major polarization axis of the alignment laser
being parallel and the minor polarization axis being perpendicular to the detector plane
($\alpha=0^{\circ}$ in main paper), whereas in the bottom row the major polarization axis is
perpendicular and the minor polarization axis is parallel to the detector plane ($\alpha=90^{\circ}$
in the main manuscript).

\section{Supplementary Note 2: \nopagebreak \\[1ex] Tomographic reconstruction of the H$^{+}$ 3D~momentum distribution}
\label{sec:tomo_rec}
\begin{figure}[bh!]
   \includegraphics{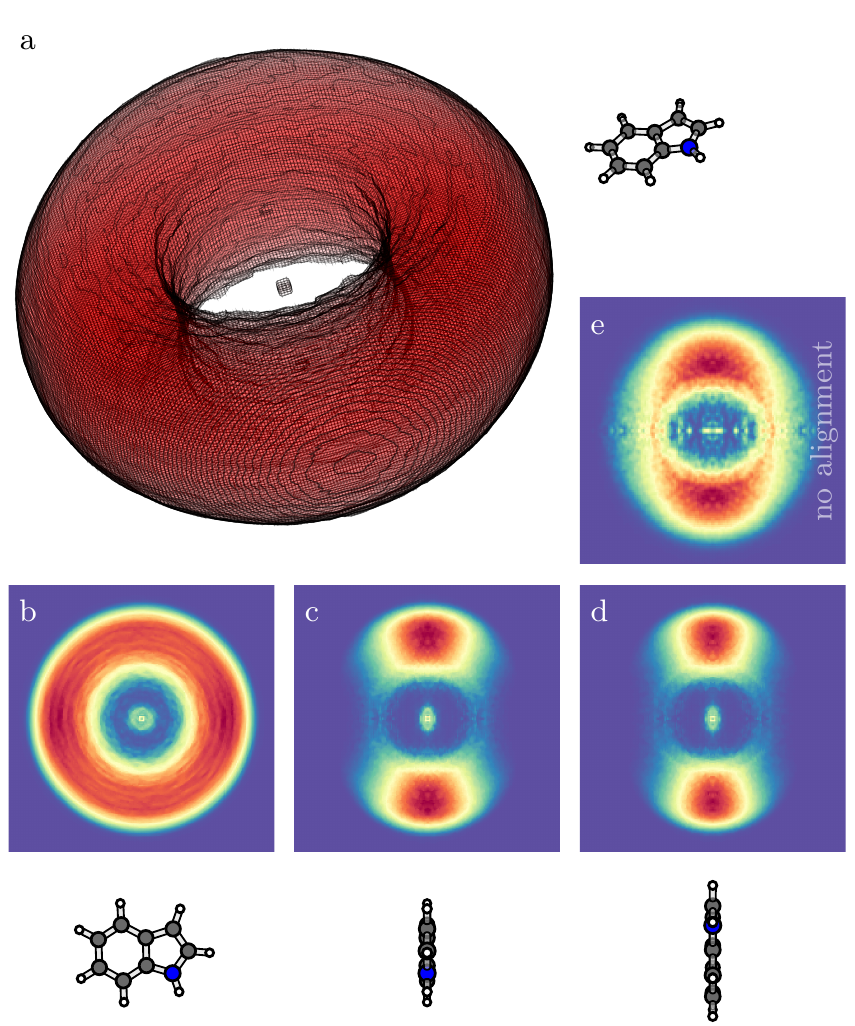}%
   \caption{\textbf{Tomographically measured and reconstructed H$^+$ ion velocity distribution},
      displayed (a) as isosurface representation and (b--d) along slices through the center of the
      distribution. The ball-and-stick models of indole depict the relative orientations of the
      molecular fixed frame. Panel (e) shows a slice through the corresponding velocity distribution
      without laser alignment; its anisotropy results from the probe selectivity of the ionization
      laser.}
   \label{fig:H_plus_tomo}
\end{figure}

The 3D momentum distribution of H$^{+}$ obtained from a tomographic reconstruction of the individual
2D projections measured at a time delay of $3.3$~ps is shown in \autoref{fig:H_plus_tomo}.
The individual 2D projections were used in the masked-VMI analysis presented in Fig.~3 of the main
article to determine the in-plane alignment distribution. Slices through the 3D ion
momentum-distribution are also shown in \autoref{fig:H_plus_tomo}. Tomographies were also acquired
at a time delay of $3.3$~ps for C$^{2+}$ and CH$_{x}$N$^{+}$($x=0,1,2$) fragments. We note that
minor orientation effects were observed due to the presence of the dc extraction field in the
interaction region of the VMI spectrometer. For the CH$_{x}$N$^{+}$($x=0,1,2$) fragments, a degree
of orientation \oricost ranging from $-0.05$ to $0.05$ was measured when the laser polarization was
rotated by \degree{90}. The 2D ion-momentum distributions of H$^{+}$ were thus symmetrized prior to
the 3D tomographic reconstruction, such that the 3D momentum distributions in
\autoref{fig:H_plus_tomo} represent the equal average over the four simultaneously present
orientations, which are related via rotations of \degree{180} around the $a$ and $b$ axes.

\section{Supplementary Note 3: \\[1ex] Computations}
\label{sec:computations}
%\subsection{Quantum Dynamics Simulations}
The rotational motion of indole was modeled in the rigid-rotor approximation using the rotational
constants $A=3877.9$~MHz, $B=1636.1$~MHz, and $C=1150.9$~MHz~\cite{Berden:JCP103:9596,
   Kang:JCP122:174301}. The electric polarizability tensor for the equilibrium molecular geometry
was computed \emph{ab initio} at the CCSD/aug-cc-pVTZ~\cite{Dunning:JCP90:1007, Kendall:JCP96:6796}
level of theory in the frozen-core approximation. Electronic-structure calculations employed the
quantum-chemistry package Dalton~\cite{Aidas:WIRE4:269}.

Time-dependent quantum dynamics simulations were performed using the general purpose code for
quantum-mechanical modelling of molecule-field interactions RichMol~\cite{Owens:JCP148:124102}. In
the simulations, the time-dependent wavefunction was built from a superposition of field-free
eigenstates with time-dependent coefficients obtained by numerically solving the time-dependent
Schrödinger equation. The latter was solved using the iterative approximation based on Krylov
subspace methods, as implemented in the Expokit computational library~\cite{Sidje:TOMS24:130}. The
elliptically polarized alignment laser field was described as
\begin{align}
  E(t) = E_0(t)\left\{\mathbf{e}_x\cos(\omega t)/\sqrt{3}, \mathbf{e}_z\sin(\omega t)\right\},
\end{align}
with $E_0(t)$, the electric field amplitude computed from the measured experimental peak intensity.
The carrier frequency was fixed to $\omega=2.354\cdot10^{15}~\text{Hz}$, corresponding to the
central wavelength $\lambda=800$~nm of the alignment laser. The time-dependent wavefunction was
expressed in the basis of field-free rotational eigenstates of indole with all rotational states
with $J\leq30$ included and propagated on a time grid with a fixed time step of 10~fs. Convergence
with respect to the size of the rotational basis set and the time step were carefully verified.

Since alignment depends nonlinearly on the laser intensity, which is not constant within the focal
volume of the laser, integration of all simulated observables over the interaction volume is
required. This has been approximated by repeating the calculations for five individual laser
intensities, obtained by scaling the originally measured peak intensity $I_0$ with factors 0.2, 0.4,
0.6, 0.8, and 1.0. Focal volume averaging was carried out using the measured Gaussian beam profiles
with widths (FWHM) of $\sigma_{\text{align}}=56.4~\um$ and $\sigma_{\text{probe}}=28.2~\um$.

\begin{figure}[bh!]
   \includegraphics[width=\linewidth]{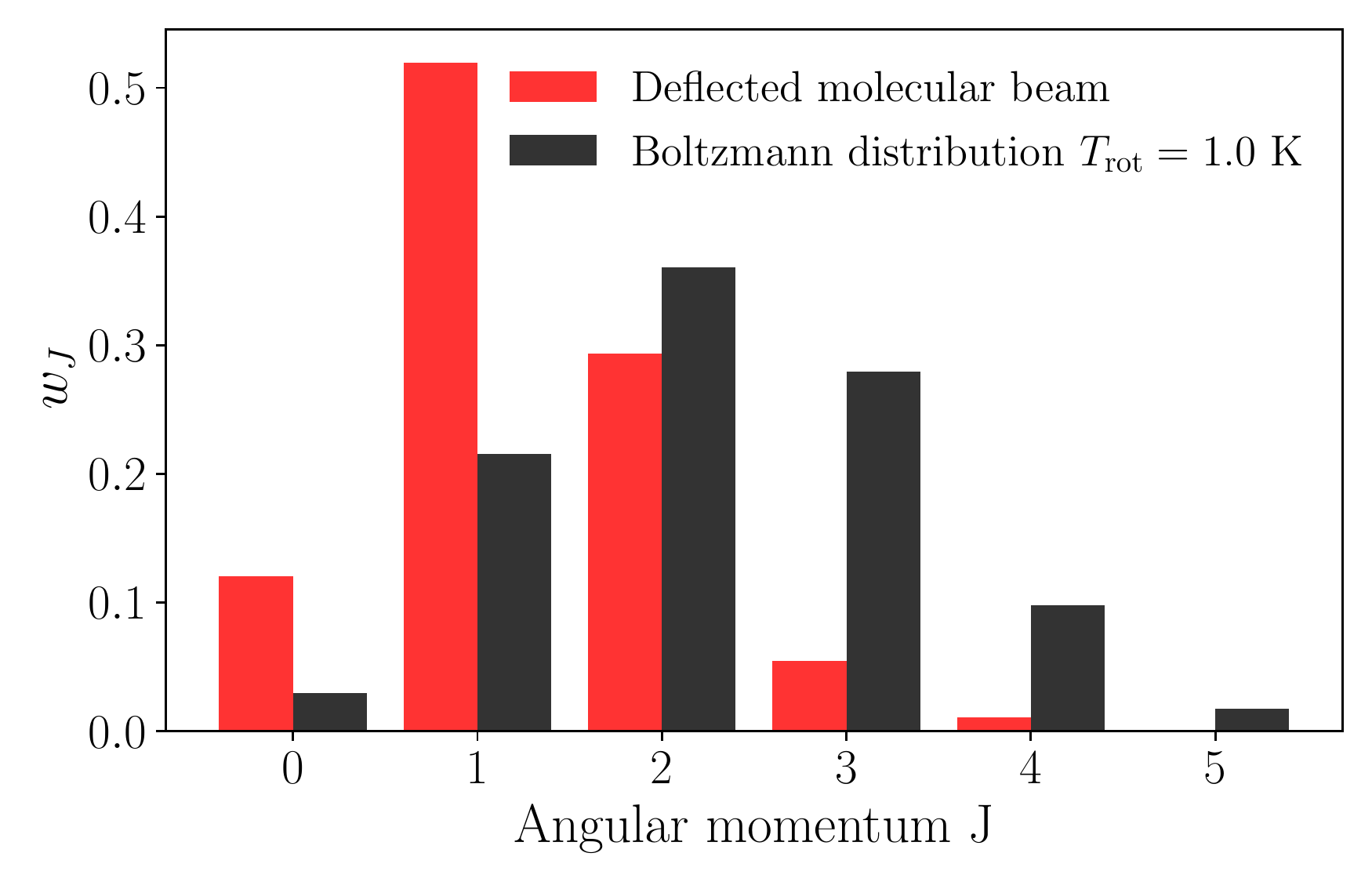}%
   \caption{\textbf{Comparison of the deflected-beam rotational-state distribution to a thermal
         Boltzmann distribution} with $T_{\text{rot}}=1.0$~K in the undeflected molecular beam. The
      weights are the sum over all sublevels with the same angular momentum quantum number $J$. The
      two distributions are clearly different. In particular, the lowest energy rotational states
      have much higher weights in the deflected beam compared to a Boltzmann distribution.}
   \label{fig:states}
\end{figure}

Finally, an incoherent average over the initial rotational-state distribution was carried out: The
rotational-state distribution of the molecule behind the deflector was determined by fitting the
measured vertical profile of the deflected molecular beam using \texttt{CMIfly}~\cite{cmifly} and is
depicted alongside a thermal distribution at 1~K in \autoref{fig:states}.
We note that the inhomogeneous electrostatic field in the deflector leads to a spatial dispersion of
rotational states according to their effective dipole moment, which is largest for the rotational
ground state. By choosing an appropriate part in the molecular beam the contribution of the
lowest-energy rotational states can be increased compared to a thermal Boltzmann distribution in the
undeflected beam~\cite{Chang:IRPC34:557}, as seen in \autoref{fig:states}.

%\subsection{Comparison of Experiment with Theory}
In order to directly compare our simulations with the experiment and to characterize the degree of
alignment, we computed the rotational probability-density distributions for all hydrogen and carbon
atoms in the molecule. The evaluation of the rotational-density functions required the calculation
of the Wigner rotation matrices, which was carried out using a Fourier-series based
algorithm~\cite{Tajima:PRC91:014320}. For different time delays, two-dimensional projections of the
rotational probability density onto the $YZ$ laboratory plane were computed for each atom
individually, assuming axial recoil of the hydrogen ions along the C-H and N-H bond vectors, and for
C$^{2+}$ the vectors connecting the center of mass with each carbon atom were chosen as recoil axes.
In analogy to the experiment, $\cost^\text{sim}_{\text{H}^+,k}$ was extracted from these 2D
projections for each hydrogen atom $k$. Furthermore, by rotating the simulated rotational
probability-density around the laboratory $Y$-axis in steps of \degree{1} and carrying out a 2D
projection for each rotation angle, the tomography measurements were mimicked. As for the
experiment, the signal within a radius of 20~pixels, calibrated to the experimental radius, was also
integrated for each angle $\alpha_{i}$ and for every hydrogen and carbon atom to reproduce our
masked-VMI measurements.

For hydrogens, the simulated time-dependent 2D alignment-revivals,
$\cost^\text{sim}_{\text{H}^+,k}$, and the angle-dependent integrated probability density,
$D_{k,\text{sim}}(\alpha_{i})$, at the center of the detector thus obtained were simultaneously
fitted to the experiment by employing a least-squares fitting routine. The fitting was achieved by
minimising the residual sum of squares, defined as:
\begin{align}
  \mathrm{RSS} =& \sum_{i=1}^{N} \left( \cost^\text{exp}_{\text{H}^+}(t_{i}) \right.\nonumber \\
  -&\left. (\sum_{k=1}^{7}w_{k}\cost^\text{sim}_{\text{H}^+,k}(t_{i})+w_{8}) \right)^2 \nonumber \\
  +&\left( \sum_{i=1}^{M}(D_{\text{exp}}(\alpha_{i}) -\sum_{k=1}^{7}w_{k}G(\alpha) \ast D_{k,\text{sim}}(\alpha_{i})) \right) ^2 \, ,
     \label{chi2}
\end{align}
with the total number $N=405$ of time steps $t_i$ and the total number $M=90$ of measured angles
$\alpha_i$ in the angle-dependent integrated probability density. $w_{k}$ and $k=1-7$ were weighting
factors associated to each hydrogen atom $k$ and $w_8$ was an offset that accounted for the
geometric alignment. $G(\alpha)\propto\exp(-\alpha^2/2w_{9}^2)$ was a Gaussian function with opening
angle $w_{9}$ (standard deviation) used to account for the non-axial recoil of the hydrogen atoms.
In the fitting procedure, a total of nine fitting parameters, $w_{k},~k=1\ldots9$, were used for a
total of 495 measured data points. Best agreement was achieved for
$\text{RSS}_\text{min}\approx0.06$ for the fitting parameters shown in
\autoref{table:fitting-parameters}. The resulting fits are shown in Fig.~2~b and Fig.~3 in the main
article, respectively. Comparison of our measured revival dynamics with the theoretical fit results
in a normalized $\chi^2$ value of 2.3, meaning, on average, our model deviates from the measured
values by approximately 1.5 times the standard error of the measured values.
\begin{table}
   \centering%
   \caption{\textbf{Parameters} obtained from the fit of the H$^+$ and C$^{2+}$ measurements shown
      in the main manuscript using the model outlined in the text.}
   \label{table:fitting-parameters}
   \begin{tabular}{c@{\qquad}c@{\qquad}c}
     \hline\hline
     Weights & H$^+$ & C$^{2+}$ \\
     \hline
     $w_1+w_3$ & $0.07 \pm 0.16$ & \\
     $w_2$ & $0.21 \pm 0.06$ & $0.07 \pm 0.01$ \\
     $w_3$ & & $0.34 \pm 0.01$ \\
     $w_4+w_7$ & $0.42 \pm 0.11$ & $0.06 \pm 0.02$\\
     $w_5+w_6$ & $0.30 \pm 0.14$ & $0.12 \pm 0.03$\\
     $w_{3a}+w_{7a}$ & & $0.41 \pm 0.02$\\
     $w_8$ & $0.081 \pm 0.021$ & \\
     $w_9$ & $50^{\circ}$ & \\
     \hline\hline
   \end{tabular}
\end{table}
A similar procedure was used to fit the angle-dependent integrated probability density measured in
the C$^{2+}$ ions. In this case, eight weighting factors, corresponding to the eight carbon atoms of
indole, were used as fitting parameters. We note that for the fit of the C$^{2+}$ ions the Gaussian
function accounting for non-axial recoil was not necessary to achieve a very good fit to the
experimental data. The parameters retrieved from the fit are also given in
\autoref{table:fitting-parameters}.

Due to averaging of the four orientations of indole the least squares fitting procedure for hydrogen
and carbon atoms with similar angles with respect to the $z_I$ axis of indole, \eg , H$_4$ \& H$_7$,
see Fig.~1 a in the main article, resulted in revivals and angle-dependent probability distributions
that are indistinguishable from each other in the experiment. This was further confirmed by
computing the covariance matrix, showing very strong correlations for near similar hydrogen and
carbon atoms, whereas no significant correlations were found between the others. The total
probabilities for these atoms were thus summed.

%\subsection{3D Degree of Alignment in the Polarizability Frame}
\begin{figure}[bh!]
   \includegraphics[width=\linewidth]{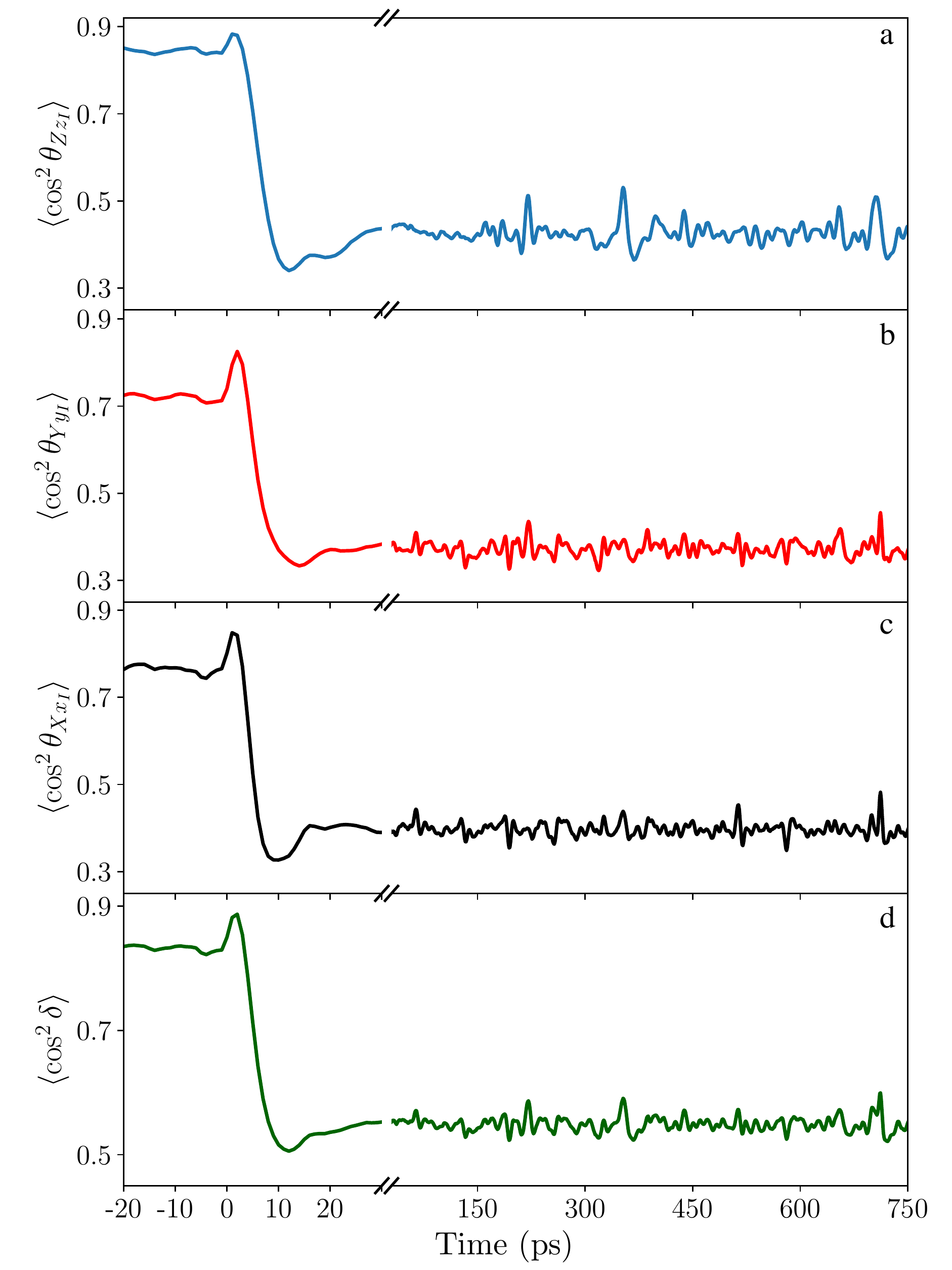}%
   \caption{\textbf{Simulated degree of 3D alignment} characterized through the expectation values
      (a) $\expectation{\cos^2\theta_{Zz_I}}$, (b) $\expectation{\cos^2\theta_{Yy_I}}$, (c)
      $\expectation{\cos^2\theta_{Xx_I}}$ and (d) $\expectation{ \cos^{2}{\delta}}$ with the
      cartesian principal axes of the polarizability tensor frame $z_{I},y_{I},x_{I}$, the cartesian
      axes of the laboratory-fixed frame $X,Y,Z$, and (d)
      $\cos^2\delta$~\cite{Makhija:PhysRevA85:033425}; see text for details.}
   \label{fig:simulated_3D_alignment}
\end{figure}
The expectation values of the alignment cosines were computed for the three main polarizability axes
in the principle-axis polarizability frame with respect to the laboratory-fixed frame by employing
Monte-Carlo integration with a convergence better than $10^{-3}$ using $\ordsim10^5$ sampling
points. The simulated degree of alignment for the main polarizability axes of the molecule
$\alpha_{z_{I}}>\alpha_{x_{I}}>\alpha_{y_{I}}$ with respect to the laboratory axes $XYZ$ is shown in
\autoref{fig:simulated_3D_alignment}. The highest achieved 3D degree of alignment was thus
characterized to be $\expectation{\cos^2\theta_{Zz_I}}=0.88$,
$\expectation{\cos^2\theta_{Yy_I}}=0.83$, $\expectation{\cos^2\theta_{Xx_I}}=0.85$, and
$\expectation{\cos^2\delta}=0.89$, where
$\cos^{2}{\delta}=\frac{1}{4}(1+\cos^2\theta_{Zz_I}+\cos^2\theta_{Yy_I}+\cos^2\theta_{Xx_I})$~\cite{Makhija:PhysRevA85:033425}.

\begin{figure}[bh!]
   \includegraphics[width=\linewidth]{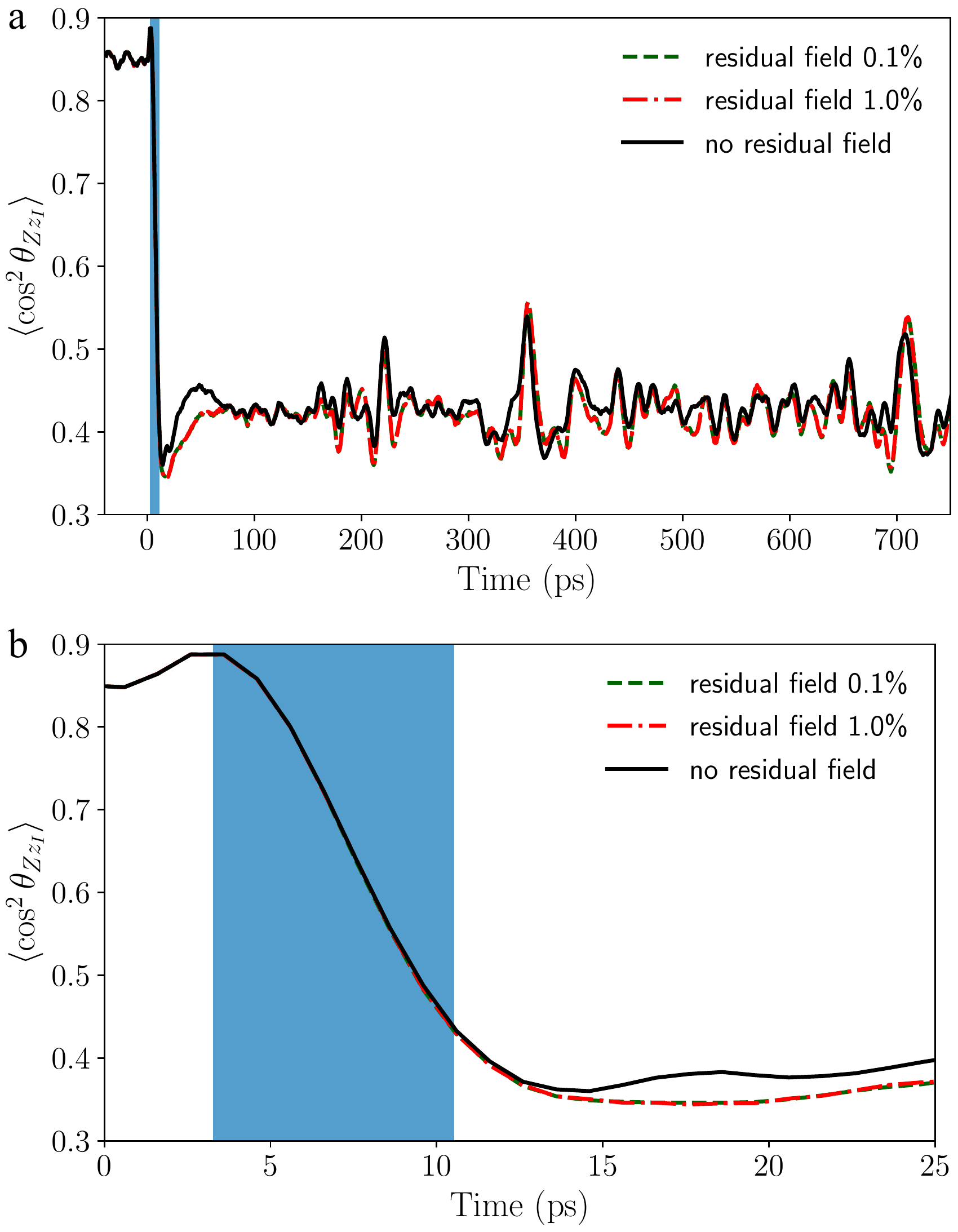}%
   \caption{\textbf{Effect of residual laser fields on the alignment.} Comparison of the 3D degree
      of alignment of the main polarizability axis $z_{I}$ with respect to the laboratory $Z$ axis
      $\expectation{\cos^{2}{\theta_{Zz_{I}}}}$ without any residual field, with a residual field of
      0.1\% of the peak intensity and with a residual field of 1\% of the peak intensity. (a) Full
      range with revivals, (b) zoom into the early times after the peak of the alignment laser
      field. In both panels the field-free region of interest is marked in blue.}
   \label{fig:residual_field}
\end{figure}
As stated in the main article, the laser field drops to below 1~\% of the peak intensity within
$3.3~\text{ps}$ after its peak value, \ie, to within the noise level of the measurement. The degree
of field-free alignment also assumes its maximum value at $3.3~\text{ps}$, which we defined as the
start of the field-free region. Simulations have been carried out considering the effect of a
residual laser field on the order of 0.1\% and 1\% of the peak intensity of the alignment laser
field and compared to the completely field-free case, shown in \autoref{fig:residual_field}. In the
simulations, the field was set to either $0.1\%$ or $1\%$ at $t=3.3$\,ps. The degree of alignment in
the region of interest, until $13~\text{ps}$ where an, generally unwanted, postpulse appears, does
not show any differences, even with such a small residual field present.

Calculations of the expected alignment pulse shape, taking into account SLM
pixelation, SLM pixel gaps, the laser beam diameter, and the spectral spread at the Fourier plane,
resulted in an expected laser intensity between $t=4$~ps and $t=10$~ps to be a factor of
$70$ lower than at the peak of the alignment pulse at $t=0$~ps.

Finally, we note that further calculations for indole (not shown) indicate that a truncation time of
$\leq2$~ps is required to obtain essentially identical dynamics to having an instantaneous
truncation. Therefore, phase shaping using the SLM based shaper was highly advantageous, or simply
necessary, instead of the more simple frequency filter used in~\cite{Chatterley:JCP148:221105},
which would result in a $8$~ps fall-off in the best case.

\renewcommand{\bibsection}{\section*{Supplementary References}}
\renewcommand{\bibsection}{\section*{References}}
\bibliography{indole-ff-alignment.bbl}
\onecolumngrid%
\listofnotes%
\end{document}